\documentclass[twocolumn,nopacs,floatfix,amsmath,nofootinbib,amssymb,preprintnumbers,floatfix]{revtex4}
\usepackage{longtable,lscape,booktabs}
\usepackage{txfonts}
\usepackage{overpic}
\usepackage{amssymb}
\usepackage{indentfirst}
\usepackage{feynmf}
\usepackage{slashed}
\usepackage{cases}
\usepackage{color}
\usepackage{multirow}
\usepackage{ulem}
\usepackage{subfigure}
\usepackage{graphicx,color,dcolumn,bm}
\usepackage{tablefootnote}
\usepackage{physics}
\usepackage{amsmath}
\usepackage{mathdots}
\usepackage{yhmath}
\usepackage{cancel}
\usepackage{color}
\usepackage{siunitx}
\usepackage{array}
\usepackage{multirow}
\usepackage{amssymb}
\usepackage{gensymb}
\usepackage{tabularx}
\usepackage{extarrows}
\usepackage{booktabs}
\usepackage{tikz}
\usetikzlibrary{fadings}
\usetikzlibrary{patterns}
\usetikzlibrary{shadows.blur}
\usetikzlibrary{shapes}
\usepackage[colorlinks,
citecolor=blue,
anchorcolor=red,
menucolor=red,
linkcolor=red,
filecolor=red,
runcolor=red,
urlcolor=blue,
frenchlinks=red]{hyperref}

\begin{document}
	
	\title{The $X(4500)$ state considered as the mixture of hadronic molecule and diquark-antidiquark within effective field theory}
	
	\author{De-Shun Zhang$^{1,2}$}
	\email[]{220220940071@lzu.edu.cn}
	\author{Wei He$^{1,2}$}
	\email[]{hewei1999@outlook.com}
	\author{Chu-Wen Xiao$^{5,6,7}$}
	\email[]{ xiaochw@gxnu.edu.cn}
	\author{Zhi-Feng Sun$^{1,2,3,4}$}
	\email[]{sunzf@lzu.edu.cn}
	
	\affiliation {\it$^1$School of Physical Science and Technology, Lanzhou University, Lanzhou 730000, China\\
		$^2$Research Center for Hadron and CSR Physics, Lanzhou University and Institute of Modern Physics of CAS, Lanzhou 730000, China\\
		$^3$Lanzhou Center for Theoretical Physics, Key Laboratory of Theoretical Physics of Gansu Province, and Key Laboratory of Quantum Theory and Applications of the Ministry of Education, Lanzhou University, Lanzhou, 730000, China\\
		$^4$Frontiers Science Center for Rare Isotopes, Lanzhou University, Lanzhou, Gansu 730000, China\\
		$^5$Department of Physics, Guangxi Normal University, Guilin 541004, China\\
		$^6$Guangxi Key Laboratory of Nuclear Physics and Technology, Guangxi Normal University, Guilin 541004, China\\
		$^7$School of Physics, Central South University, Changsha 410083, China}
	
	\date{\today}
	
	\begin{abstract}
		In the present work, we construct the Lagrangians including three-meson, meson-diquark-antidiquark vertices,  such that the diquark-antidiquark component as well as the molecular component are introduced within the effective field theory. With the obtained effective potentials projecting to spin 0, 1 and 2, we solve the Bethe-Salpeter equation with the on-shell approximation, and find that $X(4500)$ can be explained as the mixture of components $D_{s}^{*+}D_{s}^{*-}$, ${A}_{cq}\bar{A}_{cq}$ and ${A}_{cs}\bar{A}_{cs}$ with $I^G(J^{PC})=0^+(0^{++})$. In addition, another two resonances with quantum numbers $I^G(J^{PC})=0^+(1^{++})$ and $I^G(J^{PC})=0^+(2^{++})$ are predicted. 
	\end{abstract}
	
	\maketitle
	
	\section{Introduction}
	In Refs. \cite{LHCb:2016axx,LHCb:2021uow,LHCb:2024smc}, the LHCb collaboration announced their observation of four resonant states in the invariant mass spectra of $J/\psi\phi$. Two of them are named as $X(4140)$ and $X(4274)$ with $J^{PC}=1^{++}$, and the other two are named as $X(4500)$ and $X(4700)$ with $J^{PC}=0^{++}$. In the present work, we focus only on the $X(4500)$ state of which the mass and width are \cite{PDG}
	\begin{eqnarray}
		m=4474\pm4 \thinspace\text{MeV},\hspace{0.3cm} 
		\Gamma=77^{+12}_{-10}\ \text{MeV}, \nonumber
	\end{eqnarray}
	respectively.
	
	After the discovery of the LHCb collaboration mentioned above, there are many works investigating the structure of $X(4500)$. $X(4500)$ was studied by QCD sum rules in Refs. \cite{Chen:2016oma,Xie:2023xen,Wang:2016gxp,Wang:2016ujn,Wang:2021ghk}. The authors in Refs. \cite{Chen:2016oma,Xie:2023xen} explained $X(4500)$ as the D-wave tetraquark with $J^P=0^+$, while the author in Refs. \cite{Wang:2016gxp,Wang:2016ujn,Wang:2021ghk} claimed that $X(4500)$ can be explained as a 2S tetraquark with the configuration $[cs]_A[\bar{c}\bar{s}]_A$. Maiani $et\ al$ in Ref. \cite{Maiani:2016wlq} attributed $X(4140)$, $X(4500)$ and $X(4700)$ to 1S, 2S and 2S diquark-antidiquark states, respectively, and proposed the $X(4274)$ to correspond rather two, almost degenerate, unresolved lines with $J^{PC}=0^{++},2^{++}$. In the studies of the quark model \cite{Wu:2016gas,Lu:2016cwr}, $X(4500)$ is explained as the orbital or radial excitation tetraquark. However, in Ref. \cite{Yang:2019dxd} Yang and Ping found that there was no matching state to $X(4500)$ in their calculation within the chiral quark model. According to the diquark-antidiquark model \cite{Zhu:2016arf,Anwar:2018sol}, $X(4500)$ could be interpreted as the excitation of the diquark-antidiquark state in the S wave. Ref. \cite{Liu:2016onn} found that it was hard to ascribe $X(4500)$ to the rescattering effects of the P-wave threshold, which implied that $X(4500)$ could be a genuine resonace. In addition, the works \cite{Ferretti:2021xjl,Badalian:2022hfu,Badalian:2023qyi} identified the state $X(4500)$ as 4P charmonium.
	
	From the above discussion, we see that the configuration of $X(4500)$ is not so clear. Thus, in the present work we re-examine $X(4500)$ and try to reveal its nature. 
	
	In Ref. \cite{Cao:2022rjp}, we extended the hidden local symmetry to the meson-diquark sector, where the charmed mesons and charmed diquarks were viewed as the external fields, the light pseudoscalar mesons were Goldstone bosons, and the light vector mesons were gauge bosons. Then the Lagrangian of the vertices involving charmed meson, charmed diquark and light diquark were constructed. Note that only the color antitriplet diquarks were taken into account, since the interaction between the quarks of an antitriplet was attractive, while that of a color sextet was repulsive. Using the Lagrangian, we calculated the effective potentials corresponding to the mason-meson and diquark-antidiquark coupled channels. By solving the Bether-Salpeter equation, we found that the $Z_{cs}$ states could be explained as the mixture of hadronic molecule and the diquark-antidiquark state. In Ref. \cite{He:2024aej}, this formalism was extended to the bottomonium-like meson sector. By calculating the effective potentials with the Lagrangians constructed and solving the Schr\"{o}dinger equation, we found that the $Z_b$ states discovered by the Belle collaboration could be interpreted as molecular states slightly mixing with diquark-antidiquark states. 
	
	In the present work, we use the model of Refs. \cite{Cao:2022rjp,He:2024aej}, and try to study whether we could explain the $X(4500)$ state in the framework of the mixture of molecular states and diquark-antidiquark states.
	
	The structure of the paper is organized as follows. After the introduction, the theoretical framework is described in Sec. \ref{II}. The numerical results and discussion are shown in Sec. \ref{III}. Finally, a brief summary is given in Sec. \ref{IV}.
	
	\section{Formalism}\label{II}
	\subsection{The effective Lagrangian}
	In the present work, we consider the spontaneously breaking symmetry of $[U(3)_L\otimes U(3)_R]_{\text{global}}\otimes [U(3)_V]_\text{local}$, such that the pseudoscalar mesons $\pi$, $K$ and $\eta^{(\prime)}$ are introduced as Goldstone bosons, and the vectors $\rho$, $\omega$, $K^*$, $\bar{K}^*$ and $\phi$ are gauge bosons. Taking into account both the parity and charge conjugation, the Lagrangians we need are constructed as follows
	\begin{eqnarray}
		\mathcal{L}_1&=&ia_1(P\hat{\alpha}_{\|\mu} D^{\mu}P^{\dag}-D^{\mu}P\hat{\alpha}_{||\mu}P^{\dag})\nonumber\\
		&&ia_2(P\hat{\alpha}_{\bot\mu}P^{*\mu\dag}-P^{*\mu}\hat{\alpha}_{\bot\mu}P^{\dag})\nonumber\\
		&&a_3(\epsilon^{\mu\nu\alpha\beta}P^{*}_{\nu}\hat{\alpha}_{\bot\alpha}D_{\mu}P^{*\dag}_{\beta}+\epsilon^{\mu\nu\alpha\beta}D_{\mu}P^{*}_{\beta}\hat{\alpha}_{\bot\alpha}P^{*\dag}_{\nu})\nonumber\\
		&&+ia_4(P^{*}_{\nu}\hat{\alpha}_{\|}^{\mu}D_{\mu}P^{*\nu\dag}-D_{\mu}P^{*\nu}\hat{\alpha}_{\|}^{\mu}P^{*\dag}_{\nu}),\label{eq1}\\
		\mathcal{L}_2&=&e_{1}(iPD_{\mu}S^aA_{c}^{a\mu\dag}-iA_{c}^{a\mu}D_{\mu}S^{a\dag}P^{\dag})\nonumber\\
		&&+e_{2}(iPA_{\mu}^aD^{\mu}S_{c}^{a\dag}-iD^{\mu}S_{c}^aA_{\mu}^{a\dag}P^{\dag})\nonumber\\
		&&+e_{3}(\epsilon^{\mu\nu\alpha\beta}PA_{\mu\nu}^aA_{c\alpha\beta}^{a\dag}+\epsilon^{\mu\nu\alpha\beta}A_{c\alpha\beta}^aA_{\mu\nu}^{a\dag}P^{\dag})\nonumber\\
		&&+e_{4}(iP_{\mu}^{*}D^{\mu}S^aS_{c}^{a\dag}-iS_{c}^aD^{\mu}S^{a\dag}P_{\mu}^{*\dag})\nonumber\\
		&&+e_{5}(\epsilon^{\mu\nu\alpha\beta}P_{\mu}^{*}D_{\nu}S^aA_{c\alpha\beta}^{a\dag}+\epsilon^{\mu\nu\alpha\beta}A_{c\alpha\beta}^aD_{\nu}S^{a\dag}P_{\mu}^{*\dag})\nonumber\\
		&&+e_{6}(\epsilon^{\mu\nu\alpha\beta}P_{\mu}^{*}A_{\nu\alpha}^aD_{\beta}S_{c}^{a\dag}+\epsilon^{\mu\nu\alpha\beta}D_{\beta}S_{c}^aA_{\nu\alpha}^{a\dag}P_{\mu}^{*\dag})\nonumber\\
		&&+e_{7}(iP_{\mu}^{*}A^{a\mu\nu}A_{c\nu}^{a\dag}-iA_{c\nu}^aA^{a\mu\nu\dag}P_{\mu}^{*\dag})\nonumber\\
		&&+e_{8}(iP_{\mu}^{*}A_{\nu}^aA_{c}^{a\mu\nu\dag}-iA_{c}^{a\mu\nu}A_{\nu}^{a\dag}P_{\mu}^{*\dag})\nonumber\\
		&&+e_{9}(iP_{\mu\nu}^{*}A^{a\mu}A_{c}^{a\nu\dag}-iA_{c}^{a\nu}A^{a\mu\dag}P_{\mu\nu}^{*\dag}),
		\label{eq2}\\
		\mathcal{L}_3&=&h_{1}(iS_{c}^a\hat{\alpha}_{\|}^{\mu T}D_{\mu}S_{c}^{a\dag}-iD_{\mu}S_{c}^a\hat{\alpha}_{\|}^{\mu T}S_{c}^{a\dag})\nonumber\\
		&&+h_{2}(\epsilon^{\mu\nu\alpha\beta}A_{c\mu\nu}^a\hat{\alpha}_{\|\alpha}^{T}D_{\beta}S_{c}^{a\dag}+\epsilon^{\mu\nu\alpha\beta}D_{\beta}S_{c}^a\hat{\alpha}_{\|\alpha}^{T}A_{c\mu\nu}^{a\dag})\nonumber\\
		&&+h_{3}(iA_{c\mu}^a\hat{\alpha}_{\bot}^{\mu T}S_{c}^{a\dag}-iS_{c}^a\hat{\alpha}_{\bot}^{\mu T}A_{c\mu}^{a\dag})\nonumber\\
		&&+h_{4}(iA_{c\mu}^a\hat{\alpha}_{\|\nu}^{T}A_{c}^{a\mu\nu\dag}-iA_{c}^{a\mu\nu}\hat{\alpha}_{\|\nu}^{T}A_{c\mu}^{a\dag})\nonumber\\
		&&+h_{5}(\epsilon^{\mu\nu\alpha\beta}A_{c\mu}^a\hat{\alpha}_{\bot\nu}^{T}A_{c\alpha\beta}^{a\dag}+\epsilon^{\mu\nu\alpha\beta}A_{c\alpha\beta}^a\hat{\alpha}_{\bot\nu}^{T}A_{c\mu}^{a\dag}),
		\label{eq3}
	\end{eqnarray}
	where
	\begin{eqnarray}
		D_{\mu}P&=&\partial_{\mu}P+iP\alpha_{\|\mu}^{\dag}=\partial_{\mu}P+iP\alpha_{\|\mu},\nonumber\\
		D_{\mu}P^{*}_{\nu}&=&\partial_{\mu}P^{*}_{\nu}+iP^{*}_{\nu}\alpha_{\|\mu}^{\dag}=\partial_{\mu}P^{*}_{\nu}+iP^{*}_{\nu}\alpha_{\|\mu},\nonumber\\
		D_{\mu}S_{c}^{a}&=&\partial_{\mu}S_{c}^{a}-iS_{c}^{a}\alpha_{\|\mu}^{T},\nonumber\\
		D_{\mu}A^{a}_{c\nu}&=&\partial_{\mu}A^{a}_{c\nu}-iA^{a}_{c\nu}\alpha_{\|\mu}^{T},\nonumber\\
		D_{\mu}S^{a}&=&\partial_{\mu}S^{a}-iV_{\mu}S^{a}-iS^{a}V_{\mu}^{T},\nonumber\\
		D_{\mu}A^{a}_{\nu}&=&\partial_{\mu}A^{a}_{\nu}-iV_{\mu}A^{a}_{\nu}-iA^{a}_{\nu}V_{\mu}^{T},\nonumber\\
		A_{c\mu\nu}^{a}&=&D_{\mu}A_{c\nu}^{a}-D_{\nu}A_{c\mu}^{a},\nonumber\\
		A_{\mu\nu}^{a}&=&D_{\mu}A_{\nu}^{a}-D_{\nu}A_{\mu}^{a},\nonumber\\
		\alpha_{\|\mu}&=&(\partial_{\mu}\xi_{R}\xi_{R}^{\dag}+\partial_{\mu}\xi_{L}\xi_{L}^{\dag})/2i,\nonumber\\
		\alpha_{\bot\mu}&=&(\partial_{\mu}\xi_{R}\xi_{R}^{\dag}-\partial_{\mu}\xi_{L}\xi_{L}^{\dag})/2i,\nonumber\\
		\hat{\alpha}_{\|\mu}&=&\alpha_{\|\mu}-V_{\mu},\nonumber\\
		\hat{\alpha}_{\bot\mu}&=&\alpha_{\bot\mu},\nonumber\\
		\xi_{L}&=&e^{i\sigma/F_{\sigma}}e^{-i\Phi/2F_{\pi}},\nonumber\\
		\xi_{R}&=&e^{i\sigma/F_{\sigma}}e^{i\Phi/2F_{\pi}}.\nonumber
		\label{eq4}	
	\end{eqnarray}
	In the above equations, $P$ and $P^*_\mu$ are charmed meson fields, which are defined as $P=(D^0,D^+,D_s^+)$, $P^*_{\mu}=(D^{*0},D^{*+},D_s^{*+})_{\mu}$. The light mesons fields read
	\begin{eqnarray}
		\Phi&=&\sqrt{2}\left(
		\begin{array}{ccc}
			\frac{\sqrt{3}\pi^0+\eta+\sqrt{2}\eta'}{\sqrt{6}}&\pi^+&K^+\\
			\pi^-&\frac{-\sqrt{3}\pi^0+\eta+\sqrt{2}\eta'}{\sqrt{6}}&K^0\\
			K^-&\bar{K}^0&\frac{-2\eta+\sqrt{2}\eta'}{\sqrt{6}}
		\end{array}
		\right),\\
		V_{\mu}&=&\frac{g_{V}}{\sqrt{2}}\left(
		\begin{array}{ccc}
			\frac{1}{\sqrt{2}}(\rho^{0}+\omega)&\rho^+&K^{*+}\\
			\rho^-&-\frac{1}{\sqrt{2}}(\rho^{0}-\omega)&K^{*0}\\
			K^{*-}&\bar{K}^{*0}&\phi
		\end{array}
		\right)_\mu.
	\end{eqnarray}
	
	For the diquark sector, only the color antitriplet diquarks are taken into account as mentioned in Sec. I. And the color-antitriplet-diquark fields are given in the following 
	\begin{eqnarray}
		S^{a}&=&\left(
		\begin{array}{ccc}
			0&S_{ud}&S_{us}\\
			-S_{ud}&0&S_{ds}\\
			-S_{us}&-S_{ds}&0
		\end{array}
		\right)^{a},\\
		A_{\mu}^{a}&=&\left(
		\begin{array}{ccc}
			A_{uu}&\frac{1}{\sqrt{2}}A_{ud}&\frac{1}{\sqrt{2}}A_{us}\\
			\frac{1}{\sqrt{2}}A_{ud}&A_{dd}&\frac{1}{\sqrt{2}}A_{ds}\\
			\frac{1}{\sqrt{2}}A_{us}&\frac{1}{\sqrt{2}}A_{ds}&A_{ss}
		\end{array}
		\right)^{a}_{\mu},\\
		S_{c}^{a}&=&(S_{cu},S_{cd},S_{cs})^{a},\\
		A_{c\mu}^{a}&=&(A_{cu},A_{cd},A_{cs})_{\mu}^{a},
	\end{eqnarray}
	where $S^{a}$ is the light scalar diquark field, $A_{\mu}^{a}$ the light axial vector diquark field, $S_c^{a}$ the charmed scalar diquark field and $A_{c\mu}^{a}$ the charmed axial vector diquark field. The superscript $a=1,2,3$ is the color index. In Eqs. (\ref{eq2}) and (\ref{eq3}), the Einstein summation convention is used, i.e, the repeated color indexes mean the summation over them. On the other hand, the unitary gauge $\sigma=0$ is taken, such that $\xi_{L}=\xi_{R}=e^{-i\Phi/2F_{\pi}}$. 
	Comparing $\mathcal{L}_1$ with the Lagrangian in Ref. \cite{Isola:2003fh}, we have
	\begin{eqnarray}
		a_1=-\frac{\beta}{m_{P}},\ a_2=-2g,\ a_3=-\frac{g}{m_{P^*}},\ a_4=\frac{\beta}{m_{P^*}}.
	\end{eqnarray}
	For $g$, we take the full width of $D^{*+}$ from PDG \cite{PDG}, and get $g=0.58\pm 0.01$. The parameter $\beta$ is determined by the vector meson dominance \cite{Isola:2003fh,Colangelo:1993zq}, i.e., $\beta\approx 0.85$. $g_V$ is determined from the experimental value of $g_{\rho\pi\pi}$, i.e., $g_V=5.80\pm 0.91$ \cite{Harada:2003jx}. 
	For $\mathcal{L}_2$ and $\mathcal{L}_3$, there are two sets of coupling constants $e_i\ (i=1,2,...,9)$ and $h_j\ (j=1,2,...,5)$, of which the values are still unknown. In the present work, we naively use the $^3P_0$ model to determine them. Their values are listed in Table~\ref{tab:lec}. Note that we cannot fix the sign of $e_3$, $e_5$ and $e_6$, because the relative phases between the amplitudes obtained from the Lagrangian and $^3P_0$ model cannot be determined. For $e_7$, $e_8$ and $e_9$, we use the phases in Ref. \cite{Cao:2022rjp}, which explain the $Z_{cs}$ well and get the values of them. Besides, the masses of the diquarks are taken from Ref. \cite{Ferretti:2019zyh}.
	
	\begin{table}\label{I}
		\caption{The values of the low energy constants in the Lagrangian.\protect\footnotemark[1]}
		\label{tab:lec}
		\begin{tabular}{c|c|c|c|c}
			\toprule[1pt]
			$e_{1} (\text{GeV}^{-1})$&$e_{2} (\text{GeV}^{-1})$&$e_{3} (\text{GeV}^{-2})$& $e_{4} (\text{GeV}^{-1})$ &$e_{5} (\text{GeV}^{-2})$\\ 
			\hline
			-6.939&4.161&$\pm$1.520&4.039&$\pm$1.588\\
			\toprule[1pt]
			$e_{6} (\text{GeV}^{-2})$&$e_{7} (\text{GeV}^{-1})$&$e_{8} (\text{GeV}^{-1})$& $e_{9} (\text{GeV}^{-1})$& $h_{1} (\text{GeV}^{-1})$\\
			\hline
			$\pm$2.595&2.840&16.778&-11.098&-0.190 \\
			\toprule[1pt]
			$h_{2} (\text{GeV}^{-2})$&$h_{3} (\text{GeV}^{0})$& $h_{4} (\text{GeV}^{-1})$ &$h_{5} (\text{GeV}^{-1})$&\\
			\hline
			$\pm$0.356&-0.290&-1.632&$\pm$0.032&\\
			\toprule[1pt]
		\end{tabular}
	\end{table}
	
	\subsection{Flavor wave function}
	The quantum numbers of the systems that we considered in the present work are $I^{G}(J^{PC})=0^{+}(0^{++}),0^{+}(1^{++})$ and $0^{+}(2^{++})$. Recalling the isospin doublets $(A_{cu},A_{cd})$ and $(\bar{A}_{cu},-\bar{A}_{cd})$, we have the wave functions of this system 
	\begin{eqnarray}
		&|{X_{D_{s}^{*}\bar{D}_{s}^{*}}}\rangle&=|{D_{s}^{*+}D_{s}^{*-}}\rangle,\\
		&|{X_{A_{cq}\bar{A}_{cq}}}\rangle&=\frac{1}{\sqrt{2}}(|{A_{cu}\bar{A}_{cu}}\rangle+|A_{cd}\bar{A}_{cd}\rangle),\\	
		&|{X_{A_{cs}\bar{A}_{cs}}}\rangle&=|{A_{cs}\bar{A}_{cs}}\rangle.\label{eq:12}
	\end{eqnarray}
	
	\footnotetext[1]{In Ref. \cite{Cao:2022rjp}, the values of $e_7$, $e_8$ and $e_9$ in Table~\ref{tab:lec} should be $e_7=2.840$ GeV$^{-1}$, $e_8=16.778$ GeV$^{-1}$ and $e_9=-11.098$ GeV$^{-1}$. And in Ref. \cite{He:2024aej}, $h_i\ (i=1, 3, 4)$ should be corrected as $h_1=-0.084$ GeV$^{-1}$, $h_3=-0.266$ GeV$^{-1}$ and $h_4=-1.457$ GeV$^{-1}$. However, these corrections do not change the results and conclusions in these two papers.}
	
	\subsection{The effective potentials}
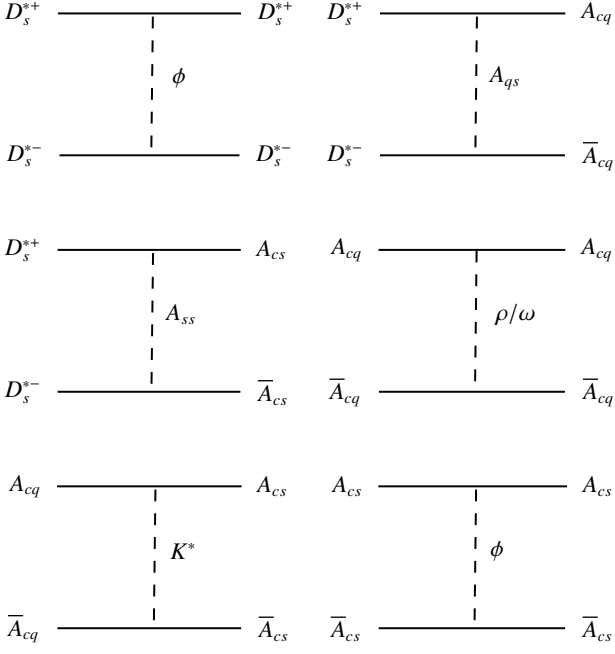
\begin{figure}
		\centering
		\tikzset{every picture/.style={line width=0.75pt}} 
		
		\begin{tikzpicture}[x=0.75pt,y=0.75pt,yscale=-1,xscale=1]
			
			\draw [color={rgb, 255:red, 0; green, 0; blue, 0 }  ,draw opacity=1 ] [dash pattern={on 4.5pt off 4.5pt}]  (95.69,19.98) -- (94.88,90.32) ;
			
			\draw [color={rgb, 255:red, 0; green, 0; blue, 0 }  ,draw opacity=1 ] [dash pattern={on 4.5pt off 4.5pt}]  (95.78,139.32) -- (94.96,209.67) ;
			\draw [color={rgb, 255:red, 0; green, 0; blue, 0 }  ,draw opacity=1 ] [dash pattern={on 4.5pt off 4.5pt}]  (259.13,19.98) -- (258.32,90.32) ;
			\draw [color={rgb, 255:red, 0; green, 0; blue, 0 }  ,draw opacity=1 ] [dash pattern={on 4.5pt off 4.5pt}]  (96.98,258.67) -- (96.17,329.01) ;
			
			\draw [color={rgb, 255:red, 0; green, 0; blue, 0 }  ,draw opacity=1 ] [dash pattern={on 4.5pt off 4.5pt}]  (258.71,258.67) -- (257.9,329.01) ;
			\draw  [dash pattern={on 4.5pt off 4.5pt}]  (258.97,138.05) -- (258.16,208.86) ;

			\draw (109.99,170.35) node    {$A_{ss}$};
			\draw (156.34,209.34) node    {$\overline{A}_{cs}$};
			\draw (30.15,209.34) node    {$D_{s}^{*-}$};
			\draw (155.53,137.73) node    {$A_{cs}$};
			\draw (30.15,137.73) node    {$D_{s}^{*+}$};
			\draw (273.35,51.01) node    {$A_{qs}$};
			\draw (320.51,89.99) node    {$\overline{A}_{cq}$};
			\draw (192.7,89.99) node    {$D_{s}^{*-}$};
			\draw (319.7,18.38) node    {$A_{cq}$};
			\draw (192.7,18.38) node    {$D_{s}^{*+}$};
			\draw (269.35,289.7) node    {$\phi $};
			\draw (320.09,328.69) node    {$\overline{A}_{cs}$};
			\draw (193.09,328.69) node    {$\overline{A}_{cs}$};
			\draw (320.09,257.08) node    {$A_{cs}$};
			\draw (193.9,257.08) node    {$A_{cs}$};
			\draw (278.64,170.35) node    {$\rho /\omega $};
			\draw (320.51,209.34) node    {$\overline{A}_{cq}$};
			\draw (192.7,209.34) node    {$\overline{A}_{cq}$};
			\draw (319.7,137.73) node    {$A_{cq}$};
			\draw (193.51,137.73) node    {$A_{cq}$};
			\draw (111.01,289.7) node    {$K^{*}$};
			\draw (156.73,328.69) node    {$\overline{A}_{cs}$};
			\draw (30.55,328.69) node    {$\overline{A}_{cq}$};
			\draw (155.92,257.08) node    {$A_{cs}$};
			\draw (31.36,257.08) node    {$A_{cq}$};
			\draw (108.86,51.01) node    {$\phi $};
			\draw (156.86,89.99) node    {$D_{s}^{*-}$};
			\draw (31.11,89.99) node    {$D_{s}^{*-}$};
			\draw (157.67,18.38) node    {$D_{s}^{*+}$};
			\draw (30.3,18.38) node    {$D_{s}^{*+}$};
			\draw    (47.65,209.34) -- (139.84,209.34) ;
			\draw    (47.65,137.73) -- (140.03,137.73) ;
			\draw    (210.2,89.99) -- (303.51,89.99) ;
			\draw    (210.2,18.38) -- (303.7,18.38) ;
			\draw    (209.59,328.69) -- (303.59,328.69) ;
			\draw    (209.4,257.08) -- (304.59,257.08) ;
			\draw    (209.7,209.34) -- (303.51,209.34) ;
			\draw    (209.51,137.73) -- (303.7,137.73) ;
			\draw    (47.55,328.69) -- (140.23,328.69) ;
			\draw    (47.36,257.08) -- (140.42,257.08) ;
			\draw    (48.61,89.99) -- (139.36,89.99) ;
			\draw    (47.8,18.38) -- (140.17,18.38) ;
			
		\end{tikzpicture}
		
		\caption{The diagrams of interaction terms on hadron levels.
		} \label{Fig1}
	\end{figure}

	The Feynman diagrams corresponding to the transitions of the channels $D_s^*\bar{D}^*_s$, $A_{cq}\bar{A}_{cq}$ and $A_{cs}\bar{A}_{cs}$ are shown in Fig. \ref{Fig1}. For the transition amplitudes obtained with the Feynman rules, we project the polarization vector products into different spin states, using the spin projectors defined as
	\begin{eqnarray}
		\mathcal{P}(0)&=&\frac{1}{3}\vec{\epsilon_{1}}\centerdot\vec{\epsilon}_{2}\vec{\epsilon}_{3}^{\dag}\centerdot\vec{\epsilon}_{4}^{\dag},\\
		\mathcal{P}(1)&=&\frac{1}{2}(\vec{\epsilon_{1}}\centerdot\vec{\epsilon}_{3}^{\dag}\vec{\epsilon}_{2}\centerdot\vec{\epsilon}_{4}^{\dag}-\vec{\epsilon_{1}}\centerdot\vec{\epsilon}_{4}^{\dag}\vec{\epsilon}_{2}\centerdot\vec{\epsilon}_{3}^{\dag}),\\
		\mathcal{P}(2)&=&\frac{1}{2}(\vec{\epsilon_{1}}\centerdot\vec{\epsilon}_{3}^{\dag}\vec{\epsilon}_{2}\centerdot\vec{\epsilon}_{4}^{\dag}+\vec{\epsilon_{1}}\centerdot\vec{\epsilon}_{4}^{\dag}\vec{\epsilon}_{2}\centerdot\vec{\epsilon}_{3}^{\dag})\nonumber\\
		&&-\frac{1}{3}\vec{\epsilon_{1}}\centerdot\vec{\epsilon}_{2}\vec{\epsilon}_{3}^{\dag}\centerdot\vec{\epsilon}_{4}^{\dag},
	\end{eqnarray}
	where $\mathcal{P}(0)$, $\mathcal{P}(1)$ and $\mathcal{P}(2)$ correspond to spin 0, 1 and 2, respectively. Finally, the effective potentials are given by
	\begin{eqnarray}
		V^{D_{s}^{*+}D_{s}^{*-}\rightarrow D_{s}^{*+}D_{s}^{*-}}_{\phi}&=&-\frac{\beta^{2}g_{V}^{2}}{2m_{P^{*}}^{2}}m_{D_{s}^{*}}^{2}(s-u)\frac{1}{m_{\phi}^{2}},\label{eq1717}\\
		V^{D_{s}^{*+}D_{s}^{*-}\rightarrow A_{cq}\bar{A}_{cq}}_{A_{qs}}&=&-\frac{\sqrt{3}}{2}m_{D_{s}^{*}}m_{A_{cq}}\left[e_{8}^{2}\frac{s-2m_{A_{cq}}^{2}}{2}+e_{8}e_{9}(m_{D_{s}^{*}}^{2}\right.\nonumber\\
		&&\left.+m_{A_{cq}}^{2}-u)+e_{9}^{2}\frac{s-2m_{D_{s}^{*}}^{2}}{2}\right]\frac{1}{m_{A_{qs}}^{2}},\label{eq18}\\
		V^{D_{s}^{*+}D_{s}^{*-}\rightarrow A_{cs}\bar{A}_{cs}}_{A_{ss}}&=&-\sqrt{3}m_{D_{s}^{*}}m_{A_{cs}}\left[e_{8}^{2}\frac{s-2m_{A_{cs}}^{2}}{2}+e_{8}e_{9}(m_{D_{s}^{*}}^{2}\right.\nonumber\\
		&&\left.+m_{A_{cs}}^{2}-u)+e_{9}^{2}\frac{s-2m_{D_{s}^{*}}^{2}}{2}\right]\frac{1}{m_{A_{ss}}^{2}},\label{eq19}
	\end{eqnarray}
	\begin{eqnarray}
		V^{A_{cq}\bar{A}_{cq}\rightarrow A_{cq}\bar{A}_{cq}}_{\rho^{0}/\omega}&=&-\frac{1}{4}h_{4}^{2}g_{V}^{2}m_{A_{cq}}^{2}(s-u)\frac{1}{m_{\rho^{0}/\omega}^{2}},\\
		V^{A_{cu}\bar{A}_{cu}\rightarrow A_{cd}\bar{A}_{cd}}_{\rho^{+}}&=&-\frac{1}{2}h_{4}^{2}g_{V}^{2}m_{A_{cu}}m_{A_{cd}}(s-u)\frac{1}{m_{\rho^{+}}^{2}},\\
		V^{A_{cu}\bar{A}_{cu}\rightarrow A_{cs}\bar{A}_{cs}}_{K^{*+}}&=&-\frac{1}{2}h_{4}^{2}g_{V}^{2}m_{A_{cu}}m_{A_{cs}}(s-u)\frac{1}{m_{K^{*+}}^{2}},\\
		V^{A_{cd}\bar{A}_{cd}\rightarrow A_{cs}\bar{A}_{cs}}_{K^{*0}}&=&-\frac{1}{2}h_{4}^{2}g_{V}^{2}m_{A_{cd}}m_{A_{cs}}(s-u)\frac{1}{m_{K^{*0}}^{2}},\\
		V^{A_{cs}\bar{A}_{cs}\rightarrow A_{cs}\bar{A}_{cs}}_{\phi}&=&-\frac{1}{2}h_{4}^{2}g_{V}^{2}m_{A_{cs}}^{2}(s-u)\frac{1}{m_{\phi}^{2}}\label{eq24}
	\end{eqnarray}
	for spin 0, 1 and 2 with the Mandelstam variables
	\begin{eqnarray}
		s&=&(p_{1}+p_{2})^{2},\\
		u&=&(p_{1}-p_{4})^{2}.
	\end{eqnarray}
	
	It is worth mentioning that the factor $\sqrt{3}$ in Eqs. \eqref{eq18} and \eqref{eq19} origins from the color wave function of the diquark-antidiquark component.\\
	
	\subsection{The Bethe-Salpeter Equation with On-shell Approximation}
	With the interaction potentials obtained above, using the Bethe-Salpeter equation in its on-shell factorized form, we obtain the scattering amplitudes of the $T$-matrix
	\begin{eqnarray}\label{eq17}
		T=(I-VG)^{-1}V,
	\end{eqnarray}
	where the matrix $V$ corresponds to the transition amplitudes, of which the elements are given by Eqs. \eqref{eq1717}-\eqref{eq24} after projecting to the s-wave. For the Mandelstam variable $u$, we neglect the product $\vec{k}_{1}\cdot\vec{k}_{3}$ which corresponds to p-wave contribution, i.e.,
	\begin{eqnarray}
		u&\approx&\frac{m_{1}^{2}+m_{2}^{2}+m_{3}^{2}+m_{4}^{2}}{2}+\frac{(m_{1}^{2}-m_{2}^{2})(m_{3}^{2}-m_{4}^{2})}{2s}-\frac{s}{2}.
		\nonumber\\
	\end{eqnarray}
	In Eq. \eqref{eq17}, $G$ is a diagonal matrix in which the nonzero element $G_{ii}$ is the two-meson loop function 
	\begin{eqnarray}\label{eq30}
		G_{ii}=i\int\frac{d^{4}q}{(2\pi)^{4}}\frac{1}{q^{2}-m^{2}_{i1}+i\epsilon}\frac{1}{(P-q)^{2}-m^{2}_{i2}+i\epsilon},
	\end{eqnarray}
	where the subscript $i$ denotes the label of a channel, $P_{\mu}$ is the total four-momentum of the two mesons in the loop, $q_{\mu}$ is the four-momentum of one of the particles, and $m_{i1}$, $m_{i2}$ are the masses of the two particles in the loop. The loop integral Eq. \eqref{eq30} is logarithmically divergent and can be regularized by the cutoff of three momentum, which has been done in Ref. \cite{Jalilian-Marian:1998tzv}, given by
	\begin{eqnarray}\label{eq20}
		G_{ii}&=&\frac{1}{32\pi^{2}}\left\{\frac{\nu}{s}\left[
		\log\frac{s-\Delta+\nu\sqrt{1+\frac{m_{i1}^{2}}{q_{max}^{2}}}}{-s+\Delta+\nu\sqrt{1+\frac{m_{i1}^{2}}{q_{max}^{2}}}}\right.\right.\nonumber\\\nonumber
		&&\left.+\log\frac{s+\Delta+\nu\sqrt{1+\frac{m_{i2}^{2}}{q_{max}^{2}}}}{-s-\Delta+\nu\sqrt{1+\frac{m_{i2}^{2}}{q_{max}^{2}}}}\right]-\frac{\Delta}{s}\log\frac{m_{i1}^{2}}{m_{i2}^{2}}\\\nonumber
		&&+\frac{2\Delta}{s}\log\frac{1+\sqrt{1+\frac{m_{i1}^{2}}{q_{max}^{2}}}}{1+\sqrt{1+\frac{m_{i2}^{2}}{q_{max}^{2}}}}+\log\frac{m_{i1}^{2}m_{i2}^{2}}{q_{max}^{4}}\\
		&&\left.-2\log\left[\left(1+\sqrt{1+\frac{m_{i1}^{2}}{q_{max}^{2}}}\right)\left(1+\sqrt{1+\frac{m_{i2}^{2}}{q_{max}^{2}}}\right)\right]\right\},\nonumber\\
		\nonumber\\
	\end{eqnarray}
	where $q_{max}$ is the three-momentum cutoff, $\Delta=m_{i2}^{2}-m_{i1}^{2}$, and $\nu=\sqrt{[s-(m_{i1}+m_{i2})^{2}][s-(m_{i1}-m_{i2})^{2}]}$
	. Eq. \eqref{eq20} justifies on the first Riemann sheet, on which the bound state locates. If one would look for the pole of resonance or virtual state, the loop function needs to be extrapolated to the second Riemann sheet, i.e.,
	\begin{eqnarray}
		G_{ii}^{II}&=&G_{ii}^{I}+i\frac{\nu}{8\pi s}.
	\end{eqnarray}
	The $T$-matrix close to a pole behaves like 
	\begin{eqnarray}
		T_{ij}&=&\frac{g_{i}g_{j}}{s-s_{R}},
	\end{eqnarray}
	where the channels $i$, $j$=$D_{s}^{*}\bar{D}_{s}^{*}$, $A_{cq}\bar{A}_{cq}$,    
	$A_{cs}\bar{A}_{cs}$, $g_{i}$ is the coupling to channel $i$, and $\text{Re}(s_{R})$ and $\text{Im}(s_{R})$ give the mass and one half of the width of the resonance found. The coupling for a certain channel is obtained as  
	\begin{eqnarray}
		g_{i}^{2}&=&\lim_{s \to s_{R}} T_{ii}(s-s_{R})
	\end{eqnarray}
	The sign of the coupling to the $D_{s}^{*}\bar{D}_{s}^{*}$ channel is chosen as positive, and those for the other channels are then determined by the following formula  
	\begin{eqnarray}
		\frac{g_{i}}{g_{j}}=\lim_{s \to s_{R}} \frac{T_{ii}}{T_{ij}}.
	\end{eqnarray}

	\section{NUMERICAL RESULTS}\label{III}
	
	\begin{table}
		\caption{Pole positions in the second Riemann sheet depending on the cutoff.}\label{tab:II}
		\begin{tabular}{c|c|c|c}
			\toprule[1pt]
			$q_{max}$ (MeV)&900&1000&1100\\
			\hline
			Pole (MeV)&4488.0$\pm$i 29.6&4485.8$\pm$i 24.9&4482.2$\pm$i 21.3\\ \bottomrule[1pt]
		\end{tabular}
	\end{table}
	
	\begin{table}
		\caption{The modules of the couplings to the $D_{s}^{*+}D_{s}^{*-}/A_{cq}\bar{A}_{cq}/A_{cs}\bar{A}_{cs}$ channels with different cutoffs.\label{tab:III}}
		\begin{tabular}{c|c|c|c}
			\toprule[1pt]
			$q_{max}$ (MeV)&900 & 1000 &1100 \\
			\hline
			$g_{D_{s}^{*+}D_{s}^{*-}}$ (MeV)& 5244 & 5408 & 5340  \\
			\hline
			$g_{A_{cq}\bar{A}_{cq}}$ (MeV)& 3784 & 3726 & 3610 \\
			\hline
			$g_{A_{cs}\bar{A}_{cs}}$ (MeV)& 10513 & 9795 & 9122 \\
			\toprule[1pt]
		\end{tabular}
	\end{table}
	
	\begin{figure}
		\centering
		\includegraphics[width=1\linewidth]{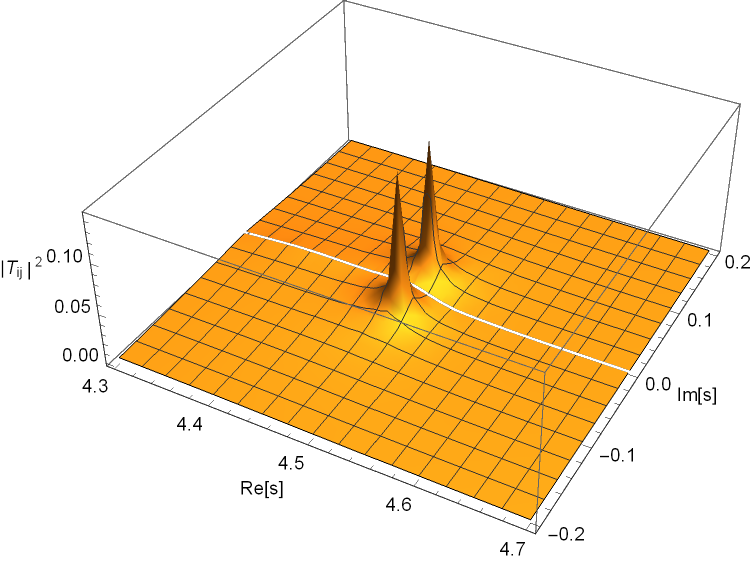}
		\caption{Three dimension plot of $|T_{ij}|^{2}$ on the second Riemann sheet with the complex variable $s$. The cutoff is chosen as 900 MeV. 
		}
		\label{fig:enter-label}
	\end{figure}
	
	In our formalism, there is only one free parameter, the cutoff in the loop functions $G_{ii}$, which is chosen as the normal value around 1 GeV with some uncertainties. By solving the Bethe-Salpeter equation, we obtain the pole positions and the modules of the couplings to all the channels depending on the cutoff, which are presented in Table \ref{tab:II} and \ref{tab:III}.
	
	For the channels $D_{s}^{*+}D_{s}^{*-}/A_{cq}\bar{A}_{cq}/A_{cs}\bar{A}_{cs}$ with quantum number $I^{G}(J^{PC})=0^{+}(0^{++})$, if reasonably choosing the cutoff $q_{max}$ as 900 MeV, we obtain a resonance pole at $(4488.0\pm29.6i)$ MeV on the second Riemann sheet. In Fig. \ref{fig:enter-label}, we plot the squared module of the $T$-matrix depending on the complex $s$. And one sees that there are a couple of poles located on the lower and upper half complex plain. This agrees with the conclusion with respect to the Schwaz reflection principle, i.e. if the pole on the lower half plain is viewed as the main pole, the one on the upper half plain is the corresponding conjugate pole. The corresponding mass and width of this pole are in good agreement with those of $X(4500)$ observed by the LHCb collaboration. In view of this result, we find that $X (4500)$ can be explained as a mixture of the channels $D_{s}^{*+}D_{s}^{*-}$, $A_{cq}\bar{A}_{cq}$, and $A_{cs}\bar{A}_{cs}$. The modules of the couplings to the three channels are obtained as $|g_{D_{s}^{*+}D_{s}^{*-}}|=5244$ MeV, $|g_{A_{cq}\bar{A}_{cq}}|=3784$ MeV, and $|g_{A_{cs}\bar{A}_{cs}}|=10513$ MeV, respectively, indicating that the $A_{cs}\bar{A}_{cs}$ component is dominant.
	
	In addition, our results predict that the resonances of the $D_{s}^{*+}D_{s}^{*-}/A_{cq}\bar{A}_{cq}/A_{cs}\bar{A}_{cs}$ systems with the quantum numbers $I^G(J^{PC})=0^+(1^{++})$ and $0^+(2^{++})$ also exist. The masses and widths of these states are the same as those of the pole found for the $X(4500)$, which are degenerate states, since the effective potentials of these two states are the same as $X(4500)$. The couplings to the channels $D_s^*\bar{D}^*_s$, $A_{cq}\bar{A}_{cq}$ and $A_{cs}\bar{A}_{cs}$ are also the same as those listed in Table \ref{tab:III}, which indicates that the $A_{cs}\bar{A}_{cs}$ component is dominant. 
	
	Finally, we simply discuss the possible decay channels of the obtained states by considering the conservation of energy, momentum, parity, and C parity. In addition to the $J/\psi\phi$ channel, these three resonances can decay into the $D^*_s\bar{D}^*_s$ and $D^*\bar{D}^*$ channels. In addition, the resonances with $I^G(J^{PC})=0^+(0^{++})$ and $0^+(1^{++})$ can also decay to the channels $D_s\bar{D}_s/D\bar{D}$ and $D_s\bar{D}^*_s/D^*_s\bar{D}_s/D\bar{D}^*/D^*\bar{D}$, respectively. Further experiments can probe the properties of these states via these decay channels.

	\section{Summary}\label{IV}
	In the present work, we employ the symmetry of ${[{U(3)}_{L}\otimes{U(3)}_{R}]}_{global}\otimes{[{U(3)}_{V}]}_{local}$ to the meson-diquark sector, and construct the Lagrangians describing the interactions of the $D_{s}^{*+}D_{s}^{*-}/A_{cq}\bar{A}_{cq}/A_{cs}\bar{A}_{cs}$ systems, where $A_{cq}\ (A_{cs})$ is the diquark component containing charm and $q\ (s)$ quark. With the effective potentials obtained by the Feynman rule, we solve the Bethe-Salpeter equation, and find a pole on the lower half-plane of the second Riemann sheet located at $(4488.0\pm i29.6)$ MeV with the cutoff $q_{max}=900$ MeV. This resonance is identified as the $X(4500)$ observed by the LHCb collaboration, since the obtained mass and width is in agreement with the experimental values measured by the LHCb collaboration. This means that $X(4500)$ can be explained as the mixture of the molecular and diquark-antidiquark components. We also evaluate the couplings to all the channels, i.e., $|g_{D_{s}^{*+}D_{s}^{*-}}|=5244$ MeV, $|g_{A_{cq}\bar{A}_{cq}}|=3784$ MeV, and $|g_{A_{cs}\bar{A}_{cs}}|=10513$ MeV, indicating that the diquark-antidiquark component $A_{cs}\bar{A}_{cs}$ is dominant. 
	
	Besides the $J/\psi \phi$ decay channel, the open charm channels, i.e., $D^{(*)}_s\bar{D}^{(*)}_s$ and $D^{(*)}\bar{D}^{(*)}$, may also be much importance for these states decaying. Future experiments can make further studies of these states via such decay channels.
	
	\section*{Acknowledgments}
	
	This project is supported by the Fundamental Research Funds for the Central Universities under Grant No. lzujbky-2022-sp02, the National Natural Science Foundation of China (NSFC) under Grants No. 11965016, 11705069 and 12335001, and the National Key Research and Development Program of China under Contract No. 2020YFA0406400, and also partly supported by the Natural Science Foundation (NSF) of Changsha under Grant No. kq2208257, the NSF of Hunan province under Grant No. 2023JJ30647, the NSF of Guangxi province under Grant No. 2023JJA110076, and the NSFC under Grant No. 12365019 (C.W.X).

	\bibliographystyle{plain}

\end{document}